\documentclass{sig-alternate-2013}

\usepackage{amsmath}
\usepackage{amsfonts}
\usepackage{graphicx}	
\usepackage{array} 

\newfont{\mycrnotice}{ptmr8t at 7pt}
\newfont{\myconfname}{ptmri8t at 7pt}
\let\crnotice\mycrnotice%
\let\confname\myconfname%

\begin{document}

\title{Poster: Revocation in VANETs Based on k-ary Huffman Trees}

\numberofauthors{1}

\author{
\alignauthor
Francisco Mart\'in-Fernandez, Pino Caballero-Gil and C\'andido Caballero-Gil\\
       \affaddr{Department of Computer Engineering}\\
       \affaddr{University of La Laguna. Tenerife}\\
       \affaddr{Spain}\\
       \email{\{fmartinf, pcaballe, ccabgil\}@ull.edu.es}
}  

\maketitle             

\begin{abstract}
Road safety has become one of the most interesting and useful topics in scientific research through the emergence of the concept  of vehicular ad-hoc network, which involves the creation of a self-managed network for interconnecting vehicles. One of the critical aspects of such networks is the need to design and implement secure and efficient control access. Thus, such a system must allow revoking malicious users, while providing access to honest users. Conventional revocation systems use simple lists to store the identification of all malicious users. Such a method is totally inefficient when the network grows and consequently the revocation list becomes unmanageable.  This paper proposes an alternative solution based on a hash tree that is used as an authenticated data structure. In particular, the solution is based on a k-ary Huffman tree, whose leaf nodes represent revoked certificates. Since vehicles that normally spend more time on the road will probably correspond to the most queried revoked certificates, this  is used to decide the corresponding depth in the tree. In this way, all the management processes related to managing  revoked certificates are optimized.

\end{abstract}

\section{Introduction}

Efficient management of  road traffic has become a  challenge in research. As a result, the so-called Vehicular Ad-hoc NETtworks (VANETs) have emerged. A VANET is a spontaneous wireless network that vehicles use for communication purpose in order to prevent hazardous events related to road traffic. Thanks to their potential, such networks can be applied to offer many different applications such as traffic information systems, intelligent freight car distribution, optimization of traffic jams or road safety.

The protection of communication security in VANETs is critical for their proper operation. Therefore, it is necessary to design an efficient system that guarantees correct authentication of network users. Thus, only legitimate users must be able to access the network,  while the access of non-legitimate and malicious users must be denied. 

Traditionally, revocation of misbehaving nodes has been solved through a centralized approach based on the existence of a Trusted Third Party (TTP), which is usually a Certificate Authority (CA) when Public Key Infrastructures (PKI) are used. The most classical solution for certificate revocation is based on the so-called Certificate Revocation Lists (CRLs), which can be seen as blacklists of revoked certificates.

This work proposes an alternative to CRLs, based on a hash tree for the efficient management of revoked certificates. Specifically, the proposal is based on a k-ary Huffman tree so that different depths are assigned depending on the time that each vehicle normally spends on the road. Regarding the hash function, the hash tree is based on a version of the standard hash function SHA-3 \cite{SHA3}.

\section{Our Proposal}

Some authors have proposed solutions to improve the computational efficiency of communications through optimized management of revocation lists in VANETs. These solutions are based on different concepts related to using Authenticated Data Structures (ADS) such as Merkle Trees \cite{hashtree} or skip lists \cite{crlADS}.
However, these proposed solutions have a high computational cost each time a new revoked certificate is added to the authenticated data structure, leading to the fact that ADS can not be constantly updated.

In order to mitigate those problems, the proposal presented here is based on k-ary Huffman trees used as hash trees, so that different depths are assigned depending on the time that each vehicle spends normally on the road. 
Specifically, the system uses hash trees to solve the scalability problem in VANETs, because large networks can be handled while preventing possible communication overheads.

The leaf nodes of the tree represent revoked certificates while nodes further up in the tree are the hashes of their respective children. The root node is the digest that represents the whole structure. The proposal uses a novel cryptographic hash function to accelerate the addition of new revoked nodes to the tree.
In particular, the proposed structure allows  inserting a new revoked certificate as a new leaf node to the right of the tree. 

In order to optimize the insertion operation, the hash function that is used is a sponge construction of the Secure Hash Algorithm SHA-3 \cite{sponge} because this allows taking advantage of the previous computation of the hash of all previous siblings without having to recompute the whole digest.
Particularly, the implemented hash function  is based on the use of a new version of  SHA-3. 

Keccak is the basic cryptographic hash function in SHA-3. Keccak function contains 24 rounds of a basic transformation and its input is represented by a $5 \times 5$ matrix of 64-bit lanes. In contrast, our proposal is based on 32-bit lanes. Another proposed variation of SHA-3 is the use of a duplex version of the sponge structure of SHA-3. The duplex construction proposal uses Keccak as fixed-length transformation, and the same padding rule and data bit rate of SHA-3. Furthermore, the duplex construction output corresponding to an input string might be obtained through the concatenation of the outputs resulting from successive input blocks, unlike a sponge function.

Another innovative idea of the presented work is related to Huffman codes, based on  building a tree where shorter paths are assigned to more frequent values. Thus, different levels are created in the tree to insert revoked certificates so that the depth of insertion into the tree corresponds to the expected popularity of the revoked certificate. Some certificates are expected to be more queried than other certificates because they correspond to vehicles that spend more time daily on the road. These vehicles, if revoked, will be located in the tree  closer to the root node.

For the initial estimation of the tree parameters, the CA uses actual data of vehicles. Thus, it asks the competent authority that owns the data in order to know the number of vehicles of each type so that it can estimate the number of levels (according to Huffman codes) and the $k$ parameter of the k-ary tree. The maximum estimated  number of revoked nodes that can have the structure is given by the results shown in   \cite{Wasef}, where it is concluded that up to 1\% of the total number of vehicles can be revoked. The size occupied by each certificate is assumed to 228 bits, according to the typical size of a classic CRL in VANET.

The TTP signature of the tree root guarantees the authenticity of the hash structure. Each time a vehicle  authenticates  another vehicle, it has to check its certificates to verify that the vehicle is legitimate and reliable in the network. The verification of certificates is performed by consulting the nearest road side unit. The response to the request must provide a  revocation proof of any revoked certificate or of that the queried certificate has not been revoked, verifiable through a signature. In case of revoked certificate, using the answer data, the vehicle can verify the TTP signature of the received signed root, recompute the root of the revocation tree, and check it by comparing it with the received signed root. 

Both the update and deletion operations of the tree are responsibility of the TTP. After each update, the TTP sends the corresponding modifications of the updated tree to all road side units. 

The proposed scheme can be considered secure because it resists many known attacks. For instance, the proposal has resistance to collusion and both backward and forward secrecy. 

This work has no weaknesses against backward secrecy attacks, because no subset of new nodes can access information sent through the network in the past. Regarding potential weaknesses to forward secrecy, the proposed scheme allows no subset of revoked nodes can access information sent through the network in the future. This is possible because the hash tree allows authentication for access control but network communication is assumed to be encrypted with session keys.

Finally, the proposal described here has resistance to collusion, thanks to the impossibility of a coalition of
two or more revoked nodes to gain access to the network by combining different available data. In fact, no coalition of revoked nodes can compromise the security of the proposal.

\section{Conclusions}

One of the biggest problems of vehicular ad-hoc networks is revocation. The efficient management of such issue has become one of the major paradigms in this area of research. A solution proposed here is based on the use of authenticated data structures like revocation trees to replace the classical and inefficient certificate revocation lists. In particular, the idea of this paper is to propose the use of k-ary hash trees, Huffman coding and a duplex version of the SHA-3 hash function, to optimize insertions and searches in the revocation structure. Thus, the inclusion of a new certificate revoked in the tree, only implies a new iteration of the duplex construction of the hash function, avoiding recalculating the entire hashes and the entire tree. Furthermore, a k-ary Huffman tree is used to insert  leaf nodes at different levels so that  those revoked nodes that are  more queried, are located closer to the root node position, so the revocation proof is smaller for those vehicles that spend more time on the roads. This paper details a method to calculate the optimum value  $k$ for the k-ary tree in order to optimize the revocation proof size. Therefore,  the proposal described here improves both the insertion of new revoked certificates in the revocation structure and the search of revoked certificates in the revocation structure. This paper is part of a work in progress, so that we plan to implement the scheme in real scenarios to get ideal values of the parameters and comparisons with other schemes.

\section{Acknowledgments} Research supported by the Spanish MINECO and the European FEDER under projects TIN2011-25452, IPT-2012-0585-370000, RTC-2014-1648-8, TEC2014-54110-R and the FPI scholarship BES-2012-051817.

\end{document}